\documentclass[11pt]{article}

\usepackage[preprint]{acl}
\usepackage{times}
\usepackage{latexsym}
\usepackage[T1]{fontenc}
\usepackage[utf8]{inputenc}
\usepackage{microtype}
\usepackage{inconsolata}
\usepackage{graphicx}

\usepackage{svg}
\usepackage{textcomp}  
\usepackage{scalerel}  
\usepackage{lipsum}
\usepackage{tikz}
\usetikzlibrary{fit,arrows,quotes,calc,automata,positioning,backgrounds,arrows.meta,bending,shapes,snakes,matrix,shadows,fadings,decorations,decorations.text,decorations.markings,shapes.geometric,shadings,tikzmark}
\usepackage{pgfplots}
\pgfplotsset{compat=1.11}
\usepackage{booktabs}
\usepackage{pgfplotstable}
\usepackage{subcaption}
\usepackage{amsmath}
\usepackage{amsfonts}
\usepackage{amssymb}
\usepackage[english]{babel}
\usepackage{mathtools}
\usepackage{hyperref}
\usepackage{mdframed}
\usepackage{csvsimple}
\usepackage{subcaption}
\usepackage{multirow}
\usepackage{tcolorbox}
\usepackage{pifont}
\usepackage{siunitx}
\usepackage{amssymb}

\definecolor{myRED}{HTML}{D50000}
\definecolor{myBLUE}{HTML}{00FBFF}
\definecolor{myDARKBLUE}{HTML}{007ACC}
\definecolor{myGREEN}{HTML}{00FF00}
\definecolor{myMAGENTA}{HTML}{C200D6}

\usepackage{soul}
\setul{0.5ex}{0.2ex}

\title{Complete Evasion, Zero Modification: PDF Attacks on AI Text Detection}

\author{Aldan Creo \\
Independent Author \\
Dublin, Ireland \\
    \href{mailto:research@acmc.fyi}{\texttt{research@acmc.fyi}} \\}

\begin{document}
\maketitle
\begin{abstract}
    AI-generated text detectors have become essential tools for maintaining content authenticity, yet their robustness against evasion attacks remains questionable. We present PDFuzz, a novel attack that exploits the discrepancy between visual text layout and extraction order in PDF documents. Our method preserves exact textual content while manipulating character positioning to scramble extraction sequences. We evaluate this approach against the ArguGPT detector using a dataset of human and AI-generated text. Our results demonstrate complete evasion: detector performance drops from \SI{93.6 +- 1.4}{\percent} accuracy and \num{0.938 +- 0.014} F1 score to random-level performance (\SI{50.4 +- 3.2}{\percent} accuracy, 0.0 F1 score) while maintaining perfect visual fidelity. Our work reveals a vulnerability in current detection systems that is inherent to PDF document structures and underscores the need for implementing sturdy safeguards against such attacks. We make our code publicly available at \url{https://github.com/ACMCMC/PDFuzz}.
\end{abstract}

\section{Introduction}\label{sec:intro}

\begin{figure*}
    \centering
    \begin{tikzpicture}[
            node distance=1.0cm,
            inner sep=0.25cm,
            font=\sffamily\small,
        ]

        \node[draw=myDARKBLUE, thick, rectangle, text width=6cm] (ai_text) {\includegraphics[width=\linewidth]{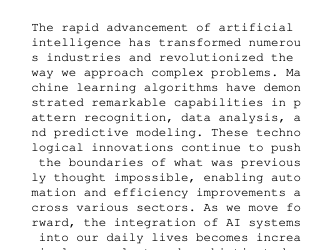}};

        \node[draw=myDARKBLUE, fill=myDARKBLUE, color=white, rectangle, below=of ai_text] (title_ai) {\textbf{Normal PDF}};

        \node[draw=myRED, thick, rectangle, text width=6cm, right=of ai_text] (attacked_text) {\includegraphics[width=\linewidth]{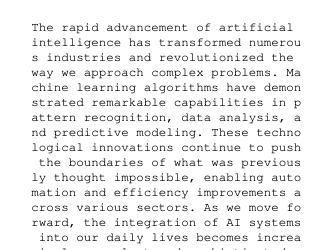}};

        \node[draw=myRED, fill=myRED, color=white, rectangle, above=of attacked_text] (title_attacked) {\textbf{Attacked PDF}};

        \node[draw=myMAGENTA, fill=myMAGENTA!10, rectangle, text width=10cm, below=of $(ai_text.south)!0.5!(attacked_text.south)$] (extractor) {{\normalsize\textbf{Extractor}}\\\vspace{0.2cm}
            Reads the encoded text in the PDF document as it is stored.};

        \node[draw=myDARKBLUE, thick, rectangle, text width=6cm, below=of ai_text |- extractor.south] (extracted_ai_text) {
            \textit{The rapid advancement of artificial\dots}
        };

        \node[draw=myRED, thick, rectangle, text width=6cm, below=of attacked_text |- extractor.south] (extracted_attacked_text) {
            \textit{Tieeimoher, nles eamwsnrlow in elnen\dots}
        };

        \draw[-, thick, myDARKBLUE] (ai_text.south) to [in=90, out=270] (extractor.north);
        \draw[-, thick, myRED] (attacked_text.south) to [in=90, out=270] (extractor.north);

        \draw[->, thick, myDARKBLUE] (extractor.south) to [in=90, out=270] (extracted_ai_text.north);
        \draw[->, thick, myRED] (extractor.south) to [in=90, out=270] (extracted_attacked_text.north);

        \node[draw=myMAGENTA, fill=myMAGENTA!10, rectangle, text width=6cm, below=2.0cm of extractor.west |- extracted_ai_text.south, anchor=north west] (detector) {{\normalsize\textbf{AI-Generated Text Detector}}\\\vspace{0.2cm}
            Classifies the text as AI-generated or not.};

        \node[draw=myDARKBLUE, thick, circle, minimum size=1cm, below=of detector.south, xshift=-2cm, inner sep=0.1cm] (ai_generated) {{\large \scalerel*{\includegraphics{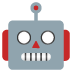}}{\textrm{\textbigcircle}}}};
        \node[draw=myRED, thick, circle, minimum size=1cm, below=of detector.south, xshift=2cm, inner sep=0.1cm] (human) {{\large \scalerel*{\includegraphics{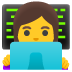}}{\textrm{\textbigcircle}}}};

        \draw[-, thick, myDARKBLUE] (extracted_ai_text.south) to [in=90, out=270] (detector.north);
        \draw[-, thick, myRED] (extracted_attacked_text.south) to [in=90, out=270] (detector.north);

        \draw[->, thick, myDARKBLUE] (detector.south) to [in=90, out=270] (ai_generated.north);
        \draw[->, thick, myRED] (detector.south) to [in=90, out=270] (human.north);

        \begin{axis}[
                at={(detector.east)},
                anchor=west,
                xshift=2cm,
                yshift=0.25cm,
                width=4cm,
                view={30}{20},
                xlabel={Step},
                zlabel={Entropy},
                xtick={0,...,9},
                ytick={1,2},
                yticklabels={\textcolor{myDARKBLUE}{{Original}},\textcolor{myRED}{{Attacked}}},
                zmin=0.0, zmax=10.0,
                grid=major,
            ]

            \coordinate (attacked_label) at (axis cs:9,2,8.15);
            \coordinate (original_label) at (axis cs:9,1,2.70);

            \addplot3[smooth, color=myDARKBLUE, mark=., opacity=0.5] coordinates {
                    (0,1,2.37) (1,1,3.81) (2,1,1.42) (3,1,2.25) (4,1,2.82) (5,1,2.16) (6,1,2.38) (7,1,3.54) (8,1,1.89) (9,1,2.23)
                };
            \addplot3[smooth, color=myDARKBLUE, mark=., opacity=0.5] coordinates {
                    (0,1,2.91) (1,1,3.12) (2,1,1.47) (3,1,2.63) (4,1,3.98) (5,1,1.22) (6,1,3.77) (7,1,3.54) (8,1,2.49) (9,1,2.43)
                };
            \addplot3[smooth, color=myDARKBLUE, mark=., opacity=0.5] coordinates {
                    (0,1,1.85) (1,1,2.67) (2,1,1.95) (3,1,2.88) (4,1,2.34) (5,1,2.71) (6,1,1.59) (7,1,2.92) (8,1,1.73) (9,1,2.11)
                };
            \addplot3[smooth, color=myDARKBLUE, mark=., opacity=0.5] coordinates {
                    (0,1,3.05) (1,1,2.44) (2,1,1.68) (3,1,2.19) (4,1,3.21) (5,1,2.03) (6,1,2.89) (7,1,3.12) (8,1,2.27) (9,1,2.56)
                };
            \addplot3[smooth, color=myDARKBLUE, mark=., opacity=0.5] coordinates {
                    (0,1,2.75) (1,1,3.10) (2,1,1.90) (3,1,2.60) (4,1,2.95) (5,1,2.40) (6,1,2.70) (7,1,3.00) (8,1,2.10) (9,1,2.80)
                };
            \addplot3[smooth, color=myDARKBLUE, mark=., thick, opacity=1.0] coordinates {
                    (0,1,2.20) (1,1,2.80) (2,1,1.60) (3,1,2.40) (4,1,3.00) (5,1,2.20) (6,1,2.50) (7,1,2.90) (8,1,2.30) (9,1,2.70)
                };

            \addplot3[smooth, color=myRED, mark=., opacity=0.5] coordinates {
                    (0,2,9.12) (1,2,8.45) (2,2,8.98) (3,2,8.21) (4,2,9.03) (5,2,8.77) (6,2,8.34) (7,2,9.28) (8,2,8.65) (9,2,8.92)
                };
            \addplot3[smooth, color=myRED, mark=., opacity=0.5] coordinates {
                    (2,2,8.11)(3,2,9.47) (4,2,5.28) (5,2,8.65) (6,2,6.02) (7,2,9.94) (8,2,7.41) (9,2,9.03)
                };
            \addplot3[smooth, color=myRED, mark=., opacity=0.5] coordinates {
                    (0,2,7.42) (1,2,5.18) (2,2,9.63) (3,2,7.07) (4,2,8.91) (5,2,6.97) (6,2,9.02) (7,2,5.15) (8,2,5.88) (9,2,5.39)
                };
            \addplot3[smooth, color=myRED, mark=., opacity=0.5] coordinates {
                    (0,2,8.12) (1,2,7.45) (2,2,8.67) (3,2,7.98) (4,2,8.21) (5,2,7.76) (6,2,8.34) (7,2,7.89) (8,2,8.23) (9,2,7.68)
                };
            \addplot3[smooth, color=myRED, mark=., opacity=0.5] coordinates {
                    (0,2,7.83) (1,2,8.12) (2,2,7.95) (3,2,8.47) (4,2,7.68) (5,2,8.21) (6,2,7.74) (7,2,8.09) (8,2,7.88) (9,2,8.15)
                };
            \addplot3[smooth, color=myRED, mark=., thick, opacity=1.0] coordinates {
                    (0,2,7.83) (1,2,8.12) (2,2,7.95) (3,2,7.37) (4,2,7.88) (5,2,7.61) (6,2,7.72) (7,2,8.01) (8,2,7.88) (9,2,8.15)
                };
        \end{axis}

        \node[anchor=north west, color=myRED, xshift=-0.5cm, yshift=1cm, inner sep=0.05cm] at (attacked_label) (note_attacked_label) {{\tiny \textit{More entropy = more ``human''}}};

        \node[anchor=north west, color=myDARKBLUE, xshift=1cm, yshift=0.6cm, inner sep=0.05cm] at (original_label) (note_original_label) {{\tiny \textit{Less entropy = more ``AI''}}};

        \draw[-, thick, myRED] (note_attacked_label.south) to [bend left=30] (attacked_label.west);
        \draw[-, thick, myDARKBLUE] (note_original_label.west) to [out=180, in=20] (original_label.west);

    \end{tikzpicture}
    \vspace{0.2cm}
    \caption{\textbf{PDFuzz.} This figure illustrates our attack methodology (as a simplified example): (1) a \textcolor{myDARKBLUE}{\textbf{Normal PDF}} contains AI-generated text with standard character ordering, appearing identical to (2) an \textcolor{myRED}{\textbf{Attacked PDF}} which uses scrambled character positioning while maintaining visual fidelity (select the texts to inspect the differences). (3) To perform AI detection, an \textcolor{myMAGENTA}{\textbf{extractor}} needs to process both documents according to their internal structure. (4) Extraction yields coherent text from the normal PDF but scrambled sequences from the attacked version. (5) The \textcolor{myMAGENTA}{\textbf{AI-generated text detector}} correctly identifies the normal text as AI-generated (\protect\scalerel*{\includegraphics{emoji_u1f916.png}}{\textrm{\textbigcircle}}) but misclassifies the scrambled text as human-written (\protect\scalerel*{\includegraphics{emoji_u1f469_200d_1f4bb.png}}{\textrm{\textbigcircle}}). The 3D plot shows perplexity patterns: \textcolor{myDARKBLUE}{original text} maintains low, stable perplexity while \textcolor{myRED}{attacked text} exhibits high, erratic perplexity due to character scrambling, causing detection failure.}
    \label{fig:approach}
\end{figure*}

The proliferation of large language models has created an urgent need for reliable AI-generated text detection systems. These detectors serve critical functions in academic integrity, content verification, and misinformation prevention. However, recent research has demonstrated that many detection systems can be circumvented through various evasion techniques.

Existing evasion methods typically involve content modification: paraphrasing attacks alter semantic meaning \cite{krishna2023paraphrasing}, character substitution techniques replace visually similar characters \cite{creo2025silverspeak}, and adversarial perturbations introduce subtle textual changes \cite{huang2024ai}. While effective, these approaches either modify the original content or introduce visually detectable artifacts. Table \ref{tab:attack-comparison} summarizes the key characteristics of existing evasion attacks and their limitations.

\begin{table*}[t]
    \centering
    \small
    \begin{tabular}{@{}p{3.5cm}cccp{6.5cm}@{}}
        \toprule
                                                             & \textbf{Same}             & \textbf{Same}             & \textbf{Same}             &                                                                                                                                                             \\
        \textbf{Attack Method}                               & \textbf{Content}          & \textbf{Appearance}       & \textbf{Meaning}          & \textbf{Main Claims}                                                                                                                                             \\
        \midrule
        DIPPER Paraphrasing \cite{krishna2023paraphrasing}   & {\color{red}\ding{55}}    & {\color{red}\ding{55}}    & {\color{red}\ding{55}}    & Their biggest drop in detection accuracy (at 1\% FPR) is from \SI{70.3}{\percent} to \SI{4.6}{\percent}; smallest from \SI{100.0}{\percent} to \SI{55.8}{\percent}. \\
        Adversarial Paraphrasing \cite{cheng2025adversarial} & {\color{red}\ding{55}}    & {\color{red}\ding{55}}    & {\color{red}\ding{55}}    & Average T@1\%F reduction of \SI{80.75}{\percent} to \SI{87.88}{\percent}, depending on the detector.                                                         \\
        Word/Sentence Substitution \cite{peng2024hidding}    & {\color{red}\ding{55}}    & {\color{red}\ding{55}}    & {\color{red}\ding{55}}    & Their perturbation methods generally reduce detection accuracy to 50-60\%.                                                                                  \\
        Token-Level Blending \cite{huang2024toblend}         & {\color{red}\ding{55}}    & {\color{red}\ding{55}}    & {\color{red}\ding{55}}    & Best result is an AUROC reduction from \num{0.9845} to \num{0.3968}, with most configurations showing modest results and some remaining unaffected.         \\
        Homoglyph Substitution \cite{creo2025silverspeak}    & {\color{red}\ding{55}}    & $\sim$                    & {\color{green}\checkmark} & Decrease the average Matthews Correlation Coefficient from \num{0.64} to \num{-0.01}.                                                                       \\
        Adversarial Perturbations \cite{zhou2024humanizing}  & {\color{red}\ding{55}}    & $\sim$                    & $\sim$                    &
        AUC dropped from \SI{99.63}{\percent} to \SI{51.06}{\percent}.                                                                                                                                                                                                                                         \\
        \midrule
        \textbf{PDFuzz (Ours)}                         & {\color{green}\checkmark} & {\color{green}\checkmark} & {\color{green}\checkmark} & \textbf{Complete evasion (F1 $\rightarrow$ 0.0)}                                                                                                            \\
        \bottomrule
    \end{tabular}
    \caption{Comparison of evasion attacks on AI text detectors. {\color{red}\ding{55}}~indicates modification/change, $\sim$~indicates minimal change, {\color{green}\checkmark}~indicates no modification/change. PDFuzz is the only method that achieves complete evasion without any content, visual, or semantic modifications.}

    \caption*{\textbf{Note:} We include main claims rather than numerical metrics because results are not directly comparable across different datasets, evaluation metrics, and experimental conditions. For complete details, we refer readers to the original articles.}
    \label{tab:attack-comparison}
\end{table*}

We identify a previously unexplored vulnerability in PDF document format: the distinction between visual character layout and text extraction order. PDF viewers display characters based on their spatial coordinates, yet text extraction follows the order in which characters are written to the document. This discrepancy creates an opportunity for evasion attacks that preserve exact textual content while scrambling extraction sequences.

Our contribution is threefold: (1) we introduce PDFuzz, the first PDF-based text ordering attack against AI detectors, (2) we demonstrate complete evasion against state-of-the-art detection systems while maintaining perfect visual fidelity, and (3) we provide empirical evidence of fundamental vulnerabilities in current detection methodologies. PDFuzz reduces the ArguGPT detector from \SI{93.6}{\percent} accuracy to essentially random performance without modifying a single character.

\section{Methods}\label{sec:methods}

\subsection{PDF Text Representation}
Unlike structured markup languages such as HTML or XML, PDF documents represent text as sequences of low-level printing commands rather than logical document structures \cite{livathinos2021robust}. Each character is positioned using absolute coordinates within a page coordinate system, where the origin (0,0) typically corresponds to the bottom-left corner of the page, with x-coordinates increasing rightward and y-coordinates increasing upward.

PDF text positioning relies on specific operators that control both character content and spatial placement \cite{iso32000}. The fundamental text-showing operators include:
\begin{itemize}
    \item[\texttt{Tj}] for displaying a string at the current text position
    \item[\texttt{TJ}] for displaying an array of strings and numeric adjustments that allow fine-tuned character spacing
    \item[\texttt{Tm}] for setting the text matrix that defines character positioning and scaling
    \item[\texttt{Td}] for moving to a new position relative to the current location
\end{itemize}
These operators enable precise control over character placement independent of reading order.

The PDF coordinate system thus allows characters to be placed at any location of choice. While conventional PDF generation writes characters sequentially from left to right and top to bottom, the specification imposes no constraints on the order in which positioning commands appear in the document stream. This flexibility enables our attack: characters can be written to the PDF in any sequence while maintaining their visual positions through explicit coordinate specification.

Text extraction tools process PDF documents by parsing the sequence of positioning commands as they appear in the document stream \cite{zhu2022pdfdataextractor}. Extraction order thus follows the order of text-showing operators in the PDF file, not the visual left-to-right, top-to-bottom reading order. This is the discrepancy that we exploit to create a gap between visual presentation and extraction sequence. In other words, while a PDF viewer makes it appear that characters are arranged in a natural reading order, we manipulate the underlying structure so that automated readers extract the text in a scrambled manner.

\subsection{Threat Model}
We assume an attacker with the following capabilities: (1) access to AI-generated text that requires evasion from detection systems and (2) ability to convert text into PDF format before submission to detectors. This represents a typical setup in most contexts of daily life, such as students submitting essays or researchers sharing manuscripts. The attacker cannot modify the original text content and must preserve visual fidelity to avoid human detection.

We assume detectors operate on extracted text sequences without format-aware preprocessing. This reflects current practice where detection systems typically receive plain text input after document parsing, making them vulnerable to extraction-order manipulation. The attack's effectiveness depends on the detection system processing text in the order defined by the PDF file structure, which should be the case for readers compliant with the PDF specification \cite{iso32000}. We do not assume access to detector internals or training data.

Our attack operates under strict constraints: no modification of character content, preservation of exact visual layout, and maintenance of document readability for human users. These constraints distinguish our approach from existing evasion methods that alter semantic content or introduce visual artifacts.

\subsection{Attack Methodology}
Our attack methodology operates in two phases. First, we analyze the target text to determine optimal character positions for normal visual presentation using monospace fonts to ensure consistent spacing. Second, we generate a modified PDF where characters are written in scrambled order while maintaining identical spatial positioning.

We implement two scrambling strategies: character-level randomization and chunk-based reordering. Character-level scrambling randomly permutes individual character positions throughout the document. Chunk-based reordering divides the text into segments of 8-15 characters and shuffles these segments while preserving intra-chunk order. Both methods preserve exact character content and visual layout.

Our implementation uses the ReportLab library to generate PDFs with precise character positioning. We calculate character coordinates based on monospace font metrics and page margins. Currently, we employ monospace fonts to simplify character alignment calculations, as each character occupies identical horizontal space. While this approach ensures reliable positioning, future implementations could incorporate sophisticated typesetting engines to support proportional fonts with varying per-glyph widths. Such enhancement would require computing individual character metrics and adjusting positions accordingly, but we leave this optimization for future work as our focus remains on demonstrating the core vulnerability rather than optimizing visual presentation.

The scrambling process maintains one-to-one character correspondence between normal and attacked versions, ensuring identical visual presentation and character counts. We verify attack success through both automated metrics comparing extracted text sequences and manual visual inspection of rendered PDFs.

\section{Results}\label{sec:results}
We evaluate our attack against the ArguGPT detector \cite{liu2023argugpt}, a RoBERTa-large model fine-tuned for identifying AI-generated argumentative essays. Our evaluation uses 1,000 essays from the SilverSpeak dataset, comprising both human-written and AI-generated content. We generate both normal PDFs and attacked versions using character-level scrambling.

The ArguGPT detector demonstrates strong baseline performance on normal text extraction, achieving \SI{93.6 +- 1.4}{\percent}\footnote{95\% confidence interval.} accuracy and \num{0.938 +- 0.014} F1 score. The detector maintains high precision at low false positive rates, with \SI{73.4}{\percent} true positive rate at \SI{1}{\percent} false positive rate.

Our PDFuzz attack reduces detector performance to random levels. Accuracy drops to \SI{50.4 +- 3.2}{\percent} and F1 score collapses to 0.0. The true positive rate at \SI{1}{\percent} false positive rate falls to \SI{0.2}{\percent}, indicating complete loss of discriminative capability.

Manual inspection of generated PDFs can confirm the identical visual presentation between normal and attacked versions (Figure \ref{fig:approach}).

\section{Discussion}\label{sec:discussion}
The attack's effectiveness stems from AI detectors' reliance on sequential text processing. Models like ArguGPT expect text to follow natural reading order, with statistical patterns reflecting normal language generation. When character extraction order becomes scrambled, these patterns are disrupted, causing detectors to misclassify the content.

This vulnerability is fundamental because it exploits the gap between human visual perception and machine text processing. While humans process text through visual pattern recognition, AI detectors analyze extracted character sequences. This creates an attack surface that preserves human-readable content while confusing machine analysis.

Potential defenses include text normalization before detection and PDF structure analysis to identify manipulation. However, these approaches face practical limitations: normalization might alter legitimate formatting, while structure analysis could be circumvented through more sophisticated positioning techniques.


\section{Conclusion}\label{sec:conclusion}
We present PDFuzz, the first PDF-based text ordering attack against AI-generated text detectors. Our method achieves complete evasion without modifying textual content, revealing fundamental vulnerabilities in current detection systems. PDFuzz demonstrates that effective evasion need not compromise visual fidelity or semantic content.

These findings highlight the need for more robust evaluation methodologies that consider document format vulnerabilities. Future detection systems should account for the distinction between visual presentation and underlying data structure across different document formats.

\section*{Limitations}\label{sec:limitations}
Our attack is specific to PDF documents and may not generalize to other formats. The effectiveness depends on the target detector's reliance on character-level sequential processing. Additionally, sophisticated detectors might implement format-aware preprocessing that could mitigate this vulnerability.

\section*{Ethics Statement}\label{sec:ethics}
Our research aims to strengthen AI detection systems by identifying vulnerabilities. We do not intend to exacerbate negative AI-related effects, such as academic misconduct or disinformation. In fact, our goal is to raise awareness of a significant vulnerability affecting current detectors. Similarly, we are making our code and techniques publicly available for academic research purposes only, and we do not allow their use for any other purpose without explicit prior authorization.

We hope our work will contribute to the development of robust detection systems which are used in a responsible manner for the greater good of society.

\bibliography{bibliography}

\begin{thebibliography}{11}
\providecommand{\natexlab}[1]{#1}

\bibitem[{Cheng et~al.(2025)Cheng, Sadasivan, Saberi, Saha, and Feizi}]{cheng2025adversarial}
Yize Cheng, Vinu~Sankar Sadasivan, Mehrdad Saberi, Shoumik Saha, and Soheil Feizi. 2025.
\newblock \href {https://arxiv.org/abs/2506.07001} {Adversarial paraphrasing: A universal attack for humanizing ai-generated text}.
\newblock \emph{Preprint}, arXiv:2506.07001.

\bibitem[{Creo and Pudasaini(2025)}]{creo2025silverspeak}
Aldan Creo and Shushanta Pudasaini. 2025.
\newblock \href {https://aclanthology.org/2025.genaidetect-1.1/} {{S}ilver{S}peak: Evading {AI}-generated text detectors using homoglyphs}.
\newblock In \emph{Proceedings of the 1stWorkshop on GenAI Content Detection (GenAIDetect)}, pages 1--46, Abu Dhabi, UAE. International Conference on Computational Linguistics.

\bibitem[{Huang et~al.(2024{\natexlab{a}})Huang, Kwak, and An}]{huang2024toblend}
Fan Huang, Haewoon Kwak, and Jisun An. 2024{\natexlab{a}}.
\newblock Toblend: Token-level blending with an ensemble of llms to attack ai-generated text detection.
\newblock \emph{arXiv preprint arXiv:2402.04776}.

\bibitem[{Huang et~al.(2024{\natexlab{b}})Huang, Zhang, Li, You, Wang, and Yang}]{huang2024ai}
Guanhua Huang, Yuchen Zhang, Zhe Li, Yongjian You, Mingze Wang, and Zhouwang Yang. 2024{\natexlab{b}}.
\newblock \href {https://doi.org/10.18653/v1/2024.acl-long.327} {Are {AI}-generated text detectors robust to adversarial perturbations?}
\newblock In \emph{Proceedings of the 62nd Annual Meeting of the Association for Computational Linguistics (Volume 1: Long Papers)}, pages 6005--6024, Bangkok, Thailand. Association for Computational Linguistics.

\bibitem[{{ISO}(2008)}]{iso32000}
{ISO}. 2008.
\newblock Document management -- portable document format -- part 1: Pdf 1.7.
\newblock Technical Report ISO 32000-1:2008, International Organization for Standardization.
\newblock Available from Adobe at \url{https://opensource.adobe.com/dc-acrobat-sdk-docs/pdfstandards/PDF32000_2008.pdf}.

\bibitem[{Krishna et~al.(2023)Krishna, Song, Karpinska, Wieting, and Iyyer}]{krishna2023paraphrasing}
Kalpesh Krishna, Yixiao Song, Marzena Karpinska, John~Frederick Wieting, and Mohit Iyyer. 2023.
\newblock \href {https://openreview.net/forum?id=WbFhFvjjKj} {Paraphrasing evades detectors of {AI}-generated text, but retrieval is an effective defense}.
\newblock In \emph{Thirty-seventh Conference on Neural Information Processing Systems}.

\bibitem[{Liu et~al.(2023)Liu, Zhang, Zhang, Yue, Zhao, Cheng, Zhang, and Hu}]{liu2023argugpt}
Yikang Liu, Ziyin Zhang, Wanyang Zhang, Shisen Yue, Xiaojing Zhao, Xinyuan Cheng, Yiwen Zhang, and Hai Hu. 2023.
\newblock Argugpt: evaluating, understanding and identifying argumentative essays generated by gpt models.
\newblock \emph{arXiv preprint arXiv:2304.07666}.

\bibitem[{Livathinos et~al.(2021)Livathinos, Berrospi, Lysak, Kuropiatnyk, Nassar, Carvalho, Dolfi, Auer, Dinkla, and Staar}]{livathinos2021robust}
Nikolaos Livathinos, Cesar Berrospi, Maksym Lysak, Viktor Kuropiatnyk, Ahmed Nassar, Andre Carvalho, Michele Dolfi, Christoph Auer, Karel Dinkla, and Peter Staar. 2021.
\newblock Robust pdf document conversion using recurrent neural networks.
\newblock In \emph{Proceedings of the AAAI Conference on Artificial Intelligence}, volume~35, pages 15137--15145.

\bibitem[{Peng et~al.(2023)Peng, Zhou, He, Sun, and Sun}]{peng2024hidding}
Xinlin Peng, Ying Zhou, Ben He, Le~Sun, and Yingfei Sun. 2023.
\newblock \href {https://openreview.net/forum?id=N4VUOeVOfS} {Hidding the ghostwriters: An adversarial evaluation of {AI}-generated student essay detection}.
\newblock In \emph{The 2023 Conference on Empirical Methods in Natural Language Processing}.

\bibitem[{Zhou et~al.(2024)Zhou, He, and Sun}]{zhou2024humanizing}
Ying Zhou, Ben He, and Le~Sun. 2024.
\newblock \href {https://aclanthology.org/2024.lrec-main.739/} {Humanizing machine-generated content: Evading {AI}-text detection through adversarial attack}.
\newblock In \emph{Proceedings of the 2024 Joint International Conference on Computational Linguistics, Language Resources and Evaluation (LREC-COLING 2024)}, pages 8427--8437, Torino, Italia. ELRA and ICCL.

\bibitem[{Zhu and Cole(2022)}]{zhu2022pdfdataextractor}
Miao Zhu and Jacqueline Cole. 2022.
\newblock Pdfdataextractor: A tool for reading scientific text and interpreting metadata from the typeset literature in the portable document format.
\newblock \emph{Journal of Chemical Information and Modeling}, 62(7):1633--1643.

\end{thebibliography}

\end{document}